\documentclass[sigplan,nonacm]{acmart}

\usepackage{placeins}
\usepackage{braket}
\usepackage{xspace}

\newcommand{\circlenumber}[1]{\raisebox{.5pt}{\textcircled{\raisebox{-.9pt} {#1}}}}
\usepackage{subcaption}  %
\usepackage{svg}
\usepackage{xcolor}
\usepackage{pagecolor}

\begin{document}
\title{ReloQate: Transient Drift Detection and In-Situ Recalibration in Surface Code Quantum Error Correction}

\author{Maxwell Poster}
\email{mposter@utexas.edu}
\orcid{0009-0006-0144-6305}
\affiliation{%
    \department{Department of Electrical and Computer Engineering}
    \institution{The University of Texas at Austin}
    \city{Austin}
    \country{USA}
}

\author{Jason Chadwick}
\email{jchadwick@uchicago.edu}
\orcid{0000-0002-7932-1418}
\affiliation{%
    \department{Department of Computer Science}
    \institution{The University of Chicago}
    \city{Chicago}
    \country{USA}
}

\author{Jonathan Mark Baker}
\email{jonathan.baker@austin.utexas.edu}
\orcid{0000-0002-0775-8274}
\affiliation{%
    \department{Department of Electrical and Computer Engineering}
    \institution{The University of Texas at Austin}
    \city{Austin}
    \country{USA}
}
\begin{abstract}
Quantum error correction (QEC) promises to exponentially suppress qubit noise, but typically assumes spatially-uniform and temporally-constant noise rates. 
However, real quantum hardware exhibits variation in noise levels over time, which will be amplified by QEC if not addressed. 
To mitigate this drift in error rates, we leverage transient information readily available in surface code quantum error correction to predict logical error rates (LER) in real time. 
We infer a prediction model by sampling physical error rates from real hardware, and mapping detector fire rate (DFR), or parity of stabilizer measurements across QEC rounds, to LER. 
This allows for on-the-fly LER predictions without the typical characterization overhead required to determine LER.
This method can easily be extended to other stabilizer codes. Importantly, we observe that this prediction should be accurate yet conservative (i.e. give an upper estimate) to enable appropriately fast responses to real-time physical error changes.
That is, responses should be executed marginally ahead of time to allow for their execution to complete, and minimize time spent (ideally none) above intolerable error rates.
More importantly, we pair this predictor with a scheme which remaps drifted logical qubits to fresh tiles in a patch-based architecture while their original tiles are recalibrated.
Our results demonstrate DFR-based prediction to be an effective LER predictor, and remapping as a spatially efficient and timely mitigation response for small code distances, both of which are significant steps in furthering practical QEC.
\end{abstract}

\maketitle %

\section{Introduction}
\begin{figure*}[!t]
    \centering
    \includegraphics[width=\textwidth,height=0.33\textheight]{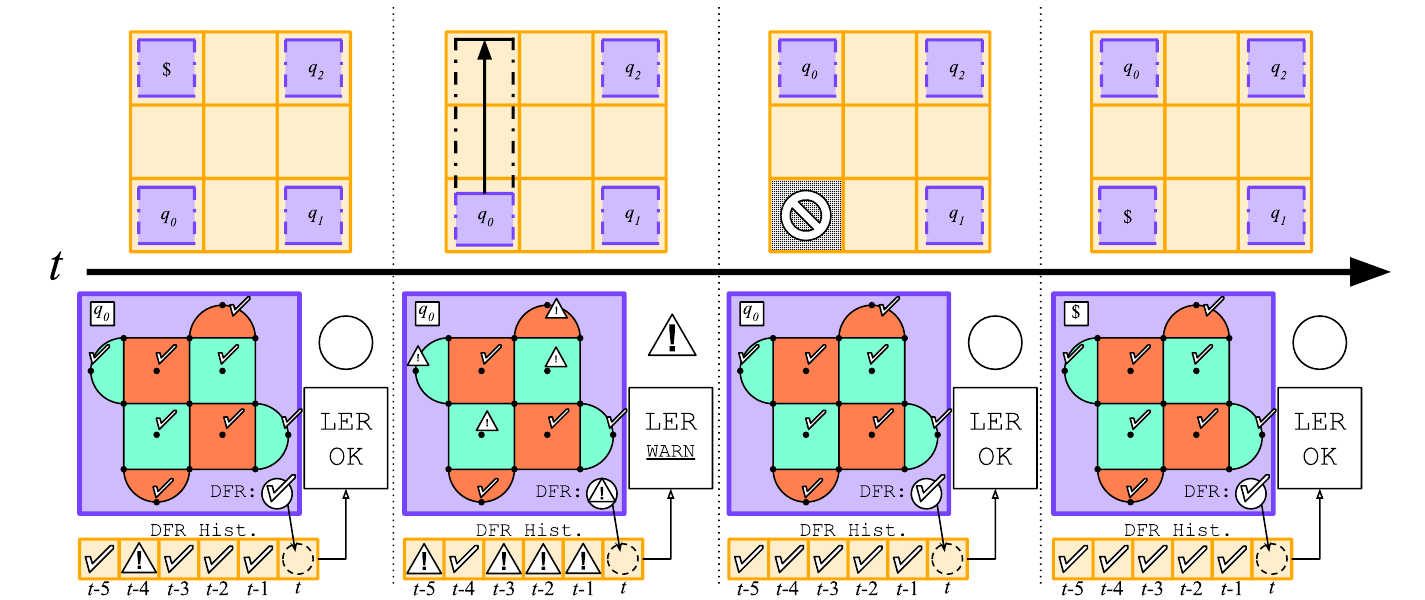}
    \caption{Timeline of a drifted qubit. During $d$ rounds of error correction, a detector fire rate (DFR) is accrued for a surface code. Detectors are fired when measurements of a parity qubit in a given round don't agree with prior measurements of that same parity qubit. The DFR is the fraction of detectors that fire (the number of fired detectors divided by the total number of detectors). Each DFR is stored within a DFR buffer as the program executes, indexed by the time it was recorded. Higher DFR is strongly correlated with an increase in LER. If multiple recent DFRs are high, it's likely the underlying physical qubits have drifted. Thus, we use the mean of all DFRs in the buffer to predict the LER of the surface code tile. If the prediction is above some target LER threshold, a remap operation is triggered. Here, $q_0$ triggers a remap operation as the mean DFR of the DFR buffer exceeded the threshold. $q_0$ is moved away to a separate tile, or reloqation target, that is below target LER. After $q_0$ has moved away, $q_0's$ original tile is disabled while it undergoes recalibration. Once complete, that tile may once again be used, and is reinitialized for use as a reloqation target.}
    \label{fig:x}
\end{figure*}

Quantum computing systems have error rates which limit the size of programs which can be successfully executed. 
One of the most promising ways to scale is to implement quantum error correction, which encodes quantum states using many physical qubits. 
Over time, physical errors still accumulate on the physical qubits which compose the logical qubit. 
Quantum error correction (QEC) operates by repeatedly measuring certain joint parities of physical qubits, which gives information related to which physical errors have occurred. If this information is accurately decoded, physical errors can be corrected before they accumulate into a logical error on the encoded state. 
While many quantum error correcting codes have been proposed, some have proven more popular than others; the surface code and various modifications are the most widely-studied group of QEC codes, because of their well-defined logical operations, low hardware connectivity requirements, and efficient decoding algorithms. 
In this work, we will focus primarily on the surface code, but the results are readily extended to other codes.

Thus far, it has been common to assume the physical error rates on component parts of the logical qubits are invariant both spatially and temporally. 
For example, a typical noise model assumes that every two qubit connection fails with a fixed probability $p$ and is the same throughout the duration of the experiment, and that this same consistency holds for single qubit error rates, decoherence, etc. 
These assumptions are useful for general benchmarking of QEC codes or decoders and are realistic for short timescales (e.g. milliseconds). 
However, on real hardware, error rates are consistent in neither space nor time. 
Several prior works have demonstrated the volatility of physical error rates in commercially available systems which can fluctuate on multiple timescales. 
For example, long-term drift can occur on the order of minutes to hours in IBM systems due to physical parameter changes such as two-level systems
\cite{meissner_probing_2018, klimov_fluctuations_2018, schlor_correlating_2019, muller_understanding_2019}. 
Others have observed catastrophic burst error mechanisms, e.g. physical error rates briefly spiking an order of magnitude due to cosmic ray events \cite{wilen_correlated_2021, mcewen_resolving_2022, thorbeck_twolevelsystem_2023, li_direct_2024}. 

Temporal inconsistency is problematic for QEC because increasing physical error rates $p$, even on a subset of the components, leads to increasingly worse logical error rates over time when the code distance remains fixed. 
The effects of variance can be mitigated most simply by increasing the code distance of the code, but at excessive resource costs. 
Similarly, systems can be \textit{engineered} to be more resilient and more consistent, for example with better calibration, control or manufacturing techniques \cite{better_engineering_2}; however, modern quantum processors still exhibit frequent drift and need regular re-calibration.

Further complicating the drift problem is the fact that, during computation or over long periods of use, the true error rates of the underlying system cannot be known exactly without performing some amount of characterization such as randomized benchmarking or tomography, both of which are infeasible to perform mid-execution. 
Consequently, during long running programs using QEC, the true error rate of each \emph{logical} qubit is also unknown. 
Fortunately, information recovered during normal execution of a logical program can be used to diagnose the existence and degree of drift occurring in the system. 
Several prior works have utilized non-intrusive metrics like syndrome weights as proxy metrics for the underlying error rates either to \circlenumber{1} modify decoding weights \cite{ken_decoding_graph_estimation,dgr, zheng_liang_2026} or \circlenumber{2} throw away low quality magic states \cite{Chadwick_2024}.
In both cases, this is insufficient for managing drift at the program scale, either because it does not allow for larger-scale drift patterns to emerge or does not apply to all qubits in the program.  
We extend these ideas of utilizing measurable quantities such as syndrome weight (also known as detector fire rate) to both detect and respond to drift at a program level. 
Independent of this work, two recent works,~\cite{zheng_liang_2026,xiao_lukin_2026}, have shown similar progress in this area, demonstrating LER prediction using syndrome data and repeated sampling. 
Unlike these, our prediction mechanism accounts for drift and works in realtime by tracking syndrome data across error correction cycles, using the program execution itself as our sampling method not requiring any additional overheads in syndrome sampling.
We draw clear correlations between detector fire rate and LER.
This correlation does not require any learning, only a simple analytical model.
Our work then shows this correlation can be used dynamically in realtime with almost no overhead (in constant time) to predict LER and inform calibration scheduling.

In this work, we propose a realtime systems architecture which dynamically adjusts program execution in response to drifting error rates. Moreover, \cite{caliscalpel, cqm} have demonstrated static approaches to the drift problem. In one \cite{cqm}, the approach requires constant movement of logical qubits to ensure consistency, but is non-dynamic. The other \cite{caliscalpel} performs in-situ recalibration of individual drifting \textit{physical qubits} in large-distance codes by deforming logical qubits around defective qubits while calibration occurs. This approach is powerful when drift rates are known ahead of time and recalibration can be statically scheduled. However, it suffers from high qubit overheads, since every logical qubit essentially needs to operate at a higher distance regularly, and, similarly to other approaches, is non-dynamic and therefore cannot be immediately responsive to more sporadic and unpredicted error patterns. Our proposal, explained in Figure \ref{fig:x}, focuses primarily on a \emph{dynamic} approach to drift detection which is complimentary to prior approaches by enabling more efficient and intelligent resource allocation without requiring any additional characterization.  

Our system has two primary components: \circlenumber{1} Detection and \circlenumber{2} Response. These components both have fairly strict operational requirements. Detection of drift should result in an accurate prediction of the logical error rate of every qubit in the system. Once the predicted logical error rate is sufficiently elevated an appropriate response can be triggered, for example by moving the logical qubit to a new location with a fresh set of recently calibrated physical qubits and labeling the old location for recalibration or by performing deformation and calibration in-place on defective qubits. Accurate prediction is critical: if the prediction is optimistic (i.e. lower than the true value) then logical qubits spend too much time in higher-error-rate locations which could lead to total program failure, while if the prediction is overly pessimistic (i.e. higher than the true value) so as to avoid accumulating too many errors, then an excess number of responses will be triggered which leads to high qubit overheads. To this end, we extensively study how to create accurate predictions which are timely yet conservative enough to eagerly avoid spiking error rates in the context of program execution.

The main contributions of this work are:
\begin{itemize}
    \item 
    We establish a clear correlation between the detector fire rate, which is transient information readily available in nominal surface code operation, and logical error rate, using data from real hardware. 
    This allows us to implement a logical error rate predictor module based on fits generated from time-varying circuits constructed with the real-world data via Stim~\cite{stim}. This predictor can be easily extended to other stabilizer codes.
    \item 
    We use these circuit simulation results to identify effective methods for tuning the performance of the predictor to be robust under both static drift models, where drift parameters may be learned, and dynamic drift models, such as those involving spontaneous burst errors. 
    \item 
    We present a drift mitigation technique which uses dedicated surface code tiles in a patch-based architecture as temporary placement for logical qubits when too large an error rate is detected. This method is efficient for small-distance surface codes, and can easily work alongside other mitigation techniques. We quantify the spatial overhead by a reloqation-qubit ratio, or the ratio between number of available relocation positions to total number of logical qubits, and identify the crossover point at which other methods may be more ideal.
    \item 
    We demonstrate remapping working in tandem with the predictor module to mitigate error drift and maintain a target logical error rate in an architectural memory experiment.
\end{itemize}

\section{Background and Prior Work}
\subsection{Quantum Error Correction with the Surface Code}
To perform fault-tolerant quantum computation (FTQC), information must be encoded into multiple physical qubits for redundancy.
The encoded information makes up a single logical qubit. The method for encoding varies based on choice of quantum error correction (QEC) code. The surface code has proven to be a reliable and practically feasible QEC code due to its convenient mapping to grid~\cite{google_qec_2024} and heavy-hex architectures~\cite{hh_map1,hh_map2}, high threshold (tolerable physical error rate) of roughly 1\%, though pays quadratic scaling in the number of required physical qubits. Alternative approaches such as qLDPC codes are gaining traction because of their high encoding rate, however, these codes are typically used primarily for memory rather than computation, where surface code is still utilized. Our approach to logical error rate detection is extensible to these other codes. In this paper, we will focus exclusively on the surface code as a good demonstration of both detection and response.
\begin{figure}[h]
    \centering
    \includegraphics[width=0.45\textwidth]{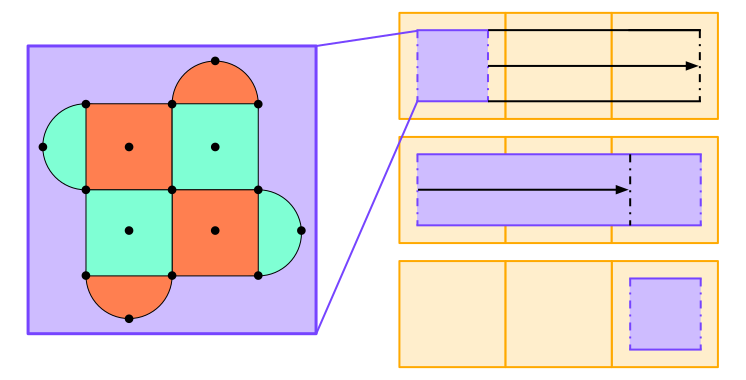}
    \caption{$d=3$ surface code undergoing a remap operation. Stored on the left-most tile of the first row is a $d=3$ surface code. Each black dot denotes a physical qubit. Black dots at intersections indicate data qubits, while black dots within solid colors or on the round edges indicate parity qubits. The surface code first expands into the far right tile in the first row. Expansion is performed such that its edge orientation (solid vs. dashed) is consistent with the original edges of the patch. In the next step, the surface code patch contracts itself (also maintaining edge consistency) by measuring away any qubits on tiles passed. The expansion operation requires a single cycle, but contraction is completed constant time (independent of cycle duration).}
    \label{fig:surface_code}
\end{figure}
The surface code is constructed from data qubits and syndrome qubits. Data qubits encode the logical information, while syndrome qubits are repeatedly measured over the execution of a program to detect errors. Syndrome qubits interact with the neighboring data qubits prior to measurement to form either a $X$- or $Z$- stabilizer, or a pauli operator of the respective stabilizer type. $X$- stabilizers detect phase-flips while $Z$- stabilizers detect bit-flips. The measurement outcomes of the stabilizers are concatenated to form a syndrome, which informs a decoder on which, if any, correction to apply to maintain the logical state of the data qubits. If the decoder cannot correct an error or proposes a correction which extends the true error into a logical operator, the code experiences a logical error. This can occur when the patch experiences a string of errors that is too long to correct. Importantly, this can also occur when the decoder's priors (knowledge of underlying physical error rates) are inaccurate, as can happen with unidentified drift.

\subsection{Lattice Surgery}
While several different surface code architectures exist, we focus on the patch-based architecture, which employs the rotated surface code in a layout of patches across a board~\cite{litinski_game}. Here, the surface code is restricted to local 2D logical operations, but has additional access to movement operations, as shown in Fig.~\ref{fig:surface_code} (or more generally, deformation). Logical qubits occupy one or more patches and may be moved arbitrary distances or reshaped with constant time overheads.

Long range interactions, in particular, are executed via lattice surgery. This involves merging the boundaries of two or more surface code patches together (possibly using extra ancillary patches to connect distance logical patches) and then later splitting the patches again. The boundary of each patch that participates in the merge determines the type of Pauli product that is measured; for example, if the $Z$ edges of two patches are merged, this performs a $ZZ$ measurement. The temporal cost of this process scales with the code distance and is independent of the spatial separation between the patches.

\subsection{Drift and calibration}
Qubits frequently experience sudden, unexpected changes to their operating parameters.
Two-level systems (TLS), thought to be caused by fabrication defects, cause unwanted interactions with the qubits, diminishing qubit coherence \cite{tls}.
Moreover, TLS frequencies are prone to shifting, so frequent calibration is required to mitigate them.
TLS are typically the strongest source of "drift" for qubits, but there exist many others.
Control fields and lab equipment may fluctuate or introduce additional noise (for example, spontaneous emission when controlling atomic qubits via laser radiation)~\cite{laserlim};
cosmic rays can introduce bursts of quasiparticles onto the chip, reducing qubit coherence locally at the site of impact (inducing TLS scrambling) for up to several hours~\cite{cosmic1,cosmic2};
and josephson junctions are prone to thermal flux~\cite{phase_slip}. 
All issues are potentially catastrophic; if unaddressed, these mechanisms ultimately limit the feasible duration of continuous computation on the device.

Calibration is frequently performed to mitigate the impact of these phenomena and maximize the likelihood of program success. Typically, the device is characterized via randomized benchmarking or gate set tomography with the goal of adjusting control parameters to minimize operational error rates. This process is both possessive, requiring full access to resources to perform calibration, and lengthy, often requiring hours to complete. This consequent monopolization of qubits leaves them unavailable for computation for long periods of time.
\begin{figure*}[!ht]
    \centering
    \includegraphics[width=0.65\textwidth]{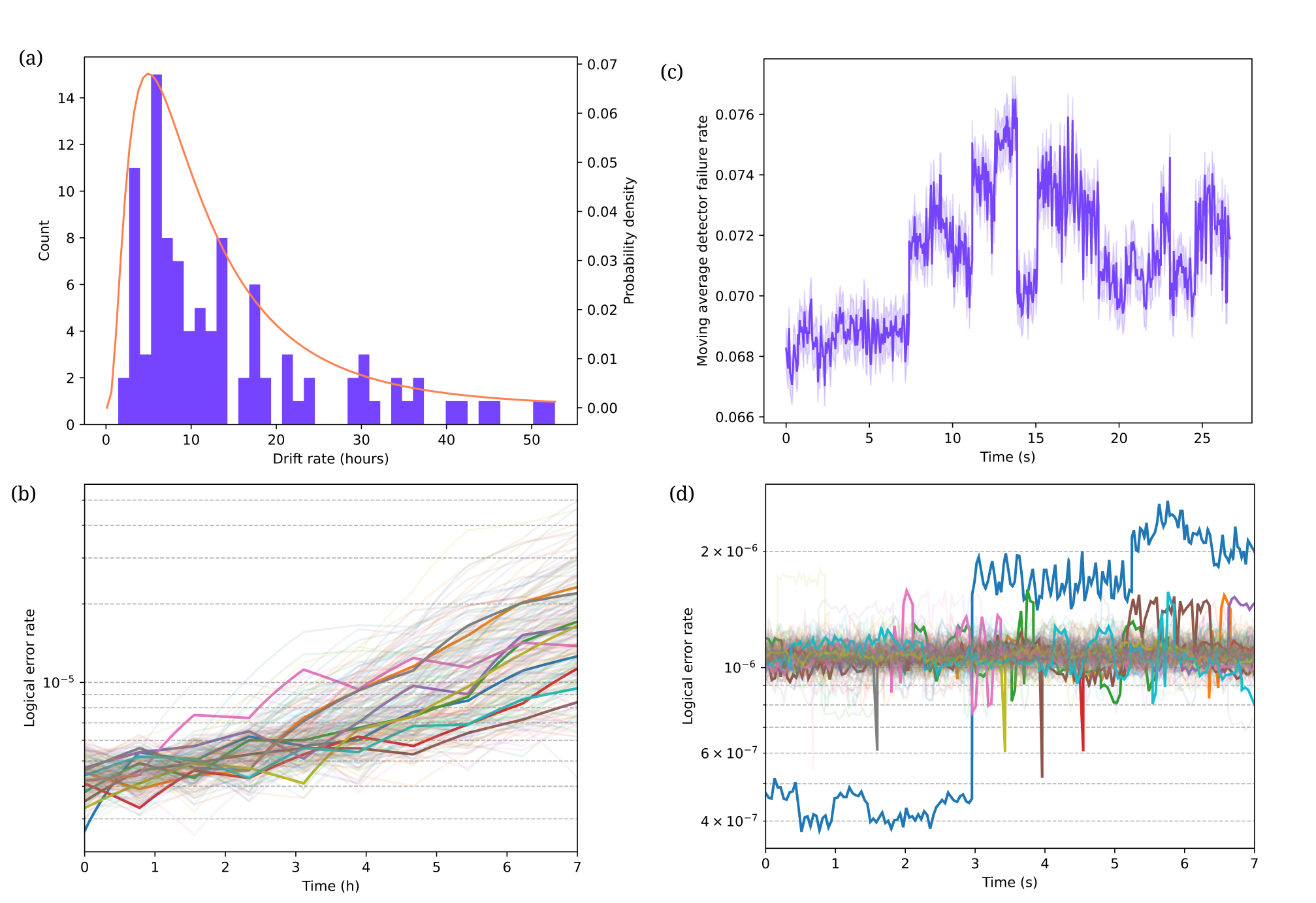}
    \caption{Examples of the drift noise models considered in this work. The \emph{slow} model exhibits drift on the timescale of hours, while the \emph{volatile} model varies on the timescale of seconds. (a) Lognormally-distributed physical qubit drift rates for the \emph{slow} model, matching the distribution observed on an IBM device in Ref. \cite{caliscalpel}. (b) Resulting logical error rate drift over time for $d=7$ surface code patches consisting of unevenly-drifting physical qubits. (c) Example trace of detector error rate over time from Google's surface code experiment \cite{google_qec_2024}. (d) Resulting logical error rate traces over time for $d=7$ surface code patches with similarly-varying detector rates.}
    \label{fig:drift-examples}
\end{figure*}

\subsection{Prior Work}

Previously proposed methods to recalibrate physical qubits mid-computation have generally involved patch deformation, which is a technique by which a surface code patch can remain operational despite a small number of ``broken'' physical qubits or couplers. These works involve deforming the surface code around quarantined qubits using superstabilizers \cite{siegel_2023, lin_codesign_2024}. While there are several different methods of performing this quarantining, all share the drawback that certain missing physical qubits or couplers require an increase in the patch size to maintain the same logical error rate. Using this method for mid-computation recalibration, as proposed in ~\cite{caliscalpel,siegel_2023, surf_deform, Q3DE}, therefore requires some amount of buffer space around each code patch to allow for expansion upon a recalibration. We instead propose shifting logical information between patches and recalibrating all physical qubits in a patch together, which significantly reduces the spatial overhead relative to patch deformation methods.

The other important component of mid-computation recalibration is fast and accurate identification of drift in the program. While it may be the case that some types of device drift occur on fixed timescales that can be learned in advance, as is assumed in Ref. \cite{caliscalpel}, we wish to be resilient to unexpected drift as well. Several prior works have suggested that the rate of error syndromes can be used to predict the overall logical performance of a patch \cite{Q3DE, Chadwick_2024}. In a similar vein, \cite{ken_decoding_graph_estimation} uses syndrome extraction to extrapolate pauli noise affecting detector error models under surface, repetition, and color codes (and finds additional suppression of LER when recalibration accounts for said noise). In this work, we develop a related approach, utilizing the correspondence between average detector fire rates and the overall logical error rate of a patch \cite{detector_likelihood}. To the best of our knowledge, our work is the first to simulate this sort of predictor on realistic device noise traces.

\section{Modeling Realtime Drift}

In this section, we introduce methods for efficiently estimating logical error rates (LERs) during long-running quantum computations using only quantities that can be measured in practice and are readily available.
To evaluate this approach, we construct LER traces that exhibit varying degrees of spatial and temporal noise variation.
These traces are then used to train a prediction model that estimates per-tile LERs using a finite window of recent syndrome measurement data.
The model is designed to be timely while also conservative, ensuring reliable estimates without overfitting to short-term fluctuations or lagging too far behind.
Finally, we describe how the system interprets sequences of predicted LERs to inform appropriate responses to anticipated drift patterns.

\subsection{Noise Models Under Drift}

It is well known~\cite{tracking_drift} that quantum hardware suffers from temporal variability in performance, typically addressed through periodic recalibration.
For short programs, this strategy is usually sufficient, as execution completes before significant drift occurs.
However, more practical, large-scale, error-corrected programs often run for significantly longer durations (e.g.
on the order of hours) making them susceptible to even slow drift over time.
For longer program traces, we model two varieties of drift in this work, \emph{slow drift} and \emph{volatile drift}. Additionally, we use a model of cosmic-ray-induced burst errors to test the response time in the most severe (and sudden) known noise scenario.

\subsubsection{Slow Drift}

In the slow drift model, each hardware component begins with a baseline physical error rate $p_0$, and this rate increases gradually over time.
Specifically, we define a drift constant: $P$ such that after $P$ seconds, the error rate has increased tenfold:
\[
p(t) = p_0 \cdot 10^{t/P}.
\]
This model is based on long-term behavior observed in IBM systems~\cite{caliscalpel,cross_variance}, where drift rates follow a lognormal distribution: $P \sim \text{Lognormal}(\mu, \sigma^2)$.
By adjusting the parameters $\mu$ and $\sigma$, we can simulate environments with varying average drift speeds and heterogeneity across components, allowing us to explore both gradual and more aggressive drift scenarios.
Examples of drift rates sampled from this model and the resulting LER traces are shown in Figure \ref{fig:drift-examples}a-b.

\subsubsection{Volatile Drift}

We utilize the extensive data accompanying Google's recent surface code experiments \cite{google_qec_2024} to create an accurate experiment-driven noise model that we call the \emph{volatile} model due to the observed sudden jumps in error rate.
The raw data consists of surface code syndrome measurements over many QEC rounds.
With knowledge of the syndrome measurement cycle time (1.1$\mu$s), we can convert these syndrome bit arrays into time-dependent DFR traces by averaging the syndrome bits in temporal windows.

We choose the window size such that the uncertainty on the calculated average is less than 10\% of the average.

We extract 35 DFR traces from the distance-5 syndrome data, with most traces typically consisting of around 30-40 seconds ($\sim$27-36 million syndrome extraction cycles).
These DFR traces can then be converted to LER traces using the fitted DFR-to-LER relationship as discussed in Section \ref{sec:dfr-vs-ler}.
Each LER trace is generated by sampling a random window of a randomly-chosen DFR trace.
Examples of a DFR trace and many randomly-generated LER traces are shown in Figure \ref{fig:drift-examples}c-d.
While most of the traces are relatively stable, some exhibit extreme LER jumps in very short times.
Being able to adapt to this volatile noise model is therefore a significant challenge.
\begin{figure*}[ht!]
    \centering
    \includegraphics[width=0.49\textwidth]{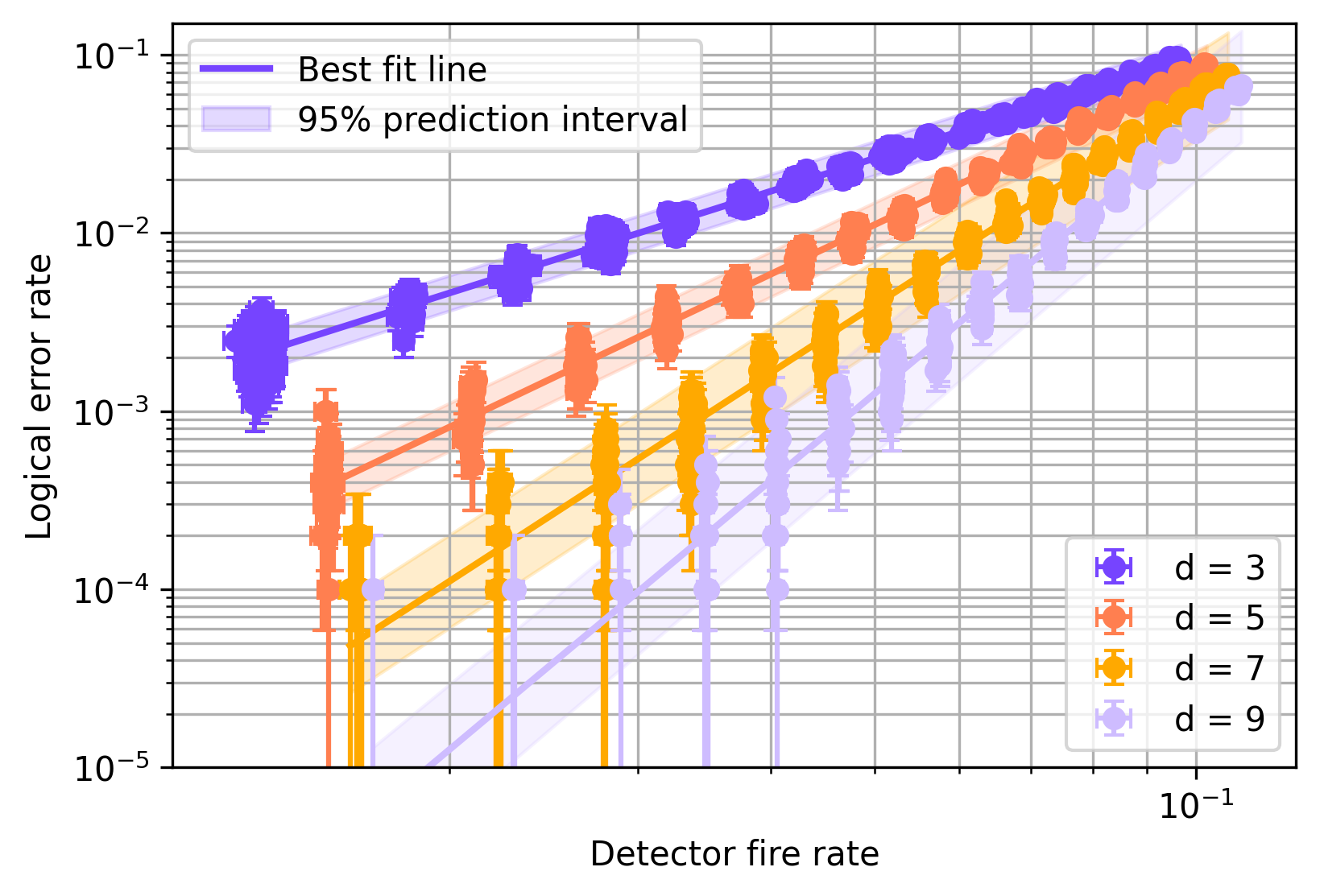}
    \includegraphics[width=0.49\textwidth]{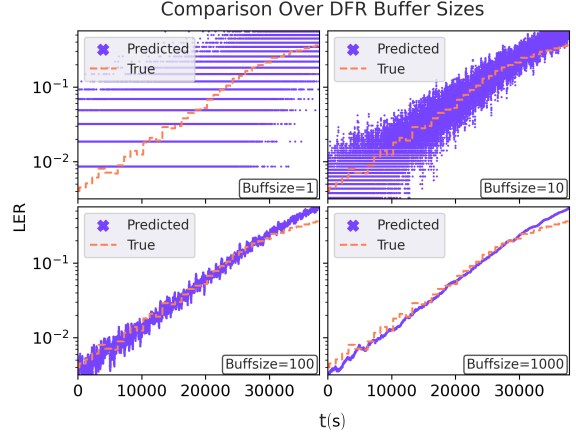}
    \caption{
        \textbf{Left:} Detector fire rates and logical error rates sampled from instances of surface codes with lognormally-distributed physical error rates. The detector fire rate of a surface code patch can be used to accurately predict the logical error rate of the patch without directly measuring it.
        \textbf{Right:} Prediction performance of an LER trace for a $d=3$ surface code under various buffer sizes over time (cycles). Larger DFR buffer sizes are generally better performing.}
    \label{fig:dfr-vs-ler}
\end{figure*}
\subsection{Logical Qubit Traces}

This work focuses on predicting logical error rates (LERs) over time under varying noise conditions.
To simulate a surface code tile, we begin by sampling the physical error rates of each component, including single-qubit gates, two-qubit gates, measurement, reset, and decoherence errors for both data and ancilla qubits-from a uniform distribution, e.g., $p_i \sim [10^{-4}, 10^{-3}]$.
This range is chosen to be below the surface code threshold, requiring non-trivial drift before rendering the tile unusable.

For continuous drift, each component is assigned a time constant sampled from a lognormal distribution.
We sample entire grids of qubits (large enough to accommodate a distance $d$ surface code) independently to capture both spatial variability in hardware and variability in qubit-to-tile mappings.
Conversely, burst drift is modeled by randomly sampling event locations, durations, and affected neighborhood sizes. We repeat this procedure independently for many tiles to capture both spatial variability in hardware and variability in qubit-to-tile mappings.

Because simulation of low physical error rates is computationally expensive (despite being polynomial-time), we cannot feasibly compute LERs at every time $t \in [0, \mathcal{T})$, where $\mathcal T$ is the total program or memory experiment duration.
Instead, we construct \textit{LER traces} by sampling at discrete time intervals and simulate the system tens to hundreds of thousands of times using Stim \cite{stim}.
These sampled traces are sufficient for our analysis, and interpolation between points approximates a continuous LER curve.

An important modeling assumption is that each measurement cycle operates under a \textit{fixed} set of physical errors.
Since the majority of drift we can respond to occurs slowly (e.g., upwards of hours) and measurement cycles are short (e.g., microseconds), error rates are effectively constant during any single cycle.
This simplifies simulation without significantly sacrificing accuracy.

For a given physical error configuration $\mathcal{E} = \{p_i\}$, each trace involves thousands of rounds of syndrome extraction.
In each round, we record the average number of fired detectors, where a detector identifies changes between sequential syndrome measurements.
If $s_i \in \{0,1\}^N$ is the syndrome in round $i$, the detector output is $s_i \oplus s_{i+1}$, the bitwise XOR of adjacent rounds.
The detector fire rate (DFR) over $r$ rounds is defined as:
\[
\text{DFR} = \sum_{i = 0}^{r-1}\frac{\sum_{j=0}^{N} (s_i \oplus s_{i+1})_j}{N}.
\]
where $N = O(d^2)$ for a distance-$d$ surface code.
Logical error rates are measured by repeatedly performing this experiment under fixed $\mathcal{E}$.
Thus, a single trace consists of many $(\text{DFR}, \text{LER})$ pairs, where LER remains constant within each segment and changes only when $\mathcal{E}$ is updated due to drift or burst events.
\subsection{Relationship Between DFR and LER Without Drift}\label{sec:dfr-vs-ler}

During the operation of a logical qubit, stabilizer measurements are the only non-intrusive quantities accessible in real time.
They thus play a central role in estimating the logical error rate (LER) of a surface code tile.
Prior (and recent) work work~\cite{detector_likelihood, xiao_lukin_2026, zheng_liang_2026, remm_wallraff_2025} has shown that the detector fire rate (DFR) serves as a good proxy for the underlying physical error rate of the system.
Since the logical error rate $p_L$ depends exponentially on the physical error rate, we expect it to scale with DFR approximately as:
\[
p_L \propto (\text{DFR})^{(d+1)/2},
\]
where $d$ is the code distance.

To test this relationship, we simulate surface code tiles with varying physical error sets $\mathcal{E} = \{p_i\}$.
Each physical error rate-including single-qubit, two-qubit, measurement, reset, and decoherence errors-is drawn from different lognormal distributions of differing mean and variance.
This allows us to study systems with low variability (e.g., trapped-ion platforms) and high variability (e.g., superconducting systems).
For each sampled model, we run thousands of syndrome extraction cycles to accurately determine the corresponding DFR and LER.

In Figure~\ref{fig:dfr-vs-ler} we see there is a strong correlation between DFR and LER across all code distances.
This indicates that, in drift-free environments, DFR can be used to reliably predict the quality of a logical qubit, providing a powerful tool for runtime monitoring since logical errors cannot be determined in realtime.
However, this correlation becomes weaker in the low-error regime.
When DFR is near zero, variance in the measured LER increases, making predictions less reliable.
This is expected; at low error rates, detectors fire infrequently over short time intervals, even though errors may still accumulate over longer durations.
This uncertainty could be mitigated by increasing the number of measurements to obtain a more accurate DFR estimate.

Doing so introduces a critical tradeoff.
The results in Figure~\ref{fig:dfr-vs-ler} assume a static noise environment, where it is safe to average DFRs over long times to reduce measurement noise.
But in the presence of drift, there are two sources of uncertainty: one from stochastic measurement noise, and one from evolving physical error rates.
If you wait too long to average out the former, you risk misattributing a real increase in errors as statistical noise which results in delayed response to a degrading logical qubit.
Therefore, while the DFR-LER correlation is highly informative in static settings, its use in dynamic environments must be approached with caution.
Accurate real-time predictions require methods that balance measurement uncertainty with responsiveness to drift.
\subsection{From Physical Drift to Logical Drift}
In order to quantify our subsequent work on dynamic detection of \textit{logical} drift, we first begin by translating the set of \emph{physical} drift constants $\{P\}$ into a tile's \emph{logical} drift constant $\tau$, and naturally quantify this as the tile's time (in seconds) to reach a $10\times$ LER.
This enables us to quantify how fast we expect the total logical error rate to change over a period of time, and so we have
\[
p_L(t) = p_L(0) \cdot 10^{t/\tau}
\]
To determine this value, we examine traces of varying drift speeds.
To produce $\tau$ we examine all of the logical errors of an input trace and fit an exponential with known value $p_L(0)$.
Faster drifting logical tiles should have lower $\tau$.

It is important to distinguish the \textit{true} LER, $p_L(t)$ (i.e.
the raw trace LER), from a \textit{predicted} LER, $\hat{p_L}(t)$ (which we will later show is derived from DFR).
In practice, $p_L$ cannot be known during execution.
Ideally, a satisfactory prediction should result in $p_L \approx \hat{p_L}$, though in practice we will actually want $p_L < \hat{p_L}$ in an effort to be zealous in any response if required.
As we will now see, the quality of this time-dependent prediction depends on the window of DFR information we use.
The faster the drift, the smaller the window we should use to capture more short-term fluctuations while still being conservative.

\section{Detecting and Predicting Temporal Fluctuations}
\begin{figure}[!h]
    \centering
    \includegraphics[width=0.45\textwidth]{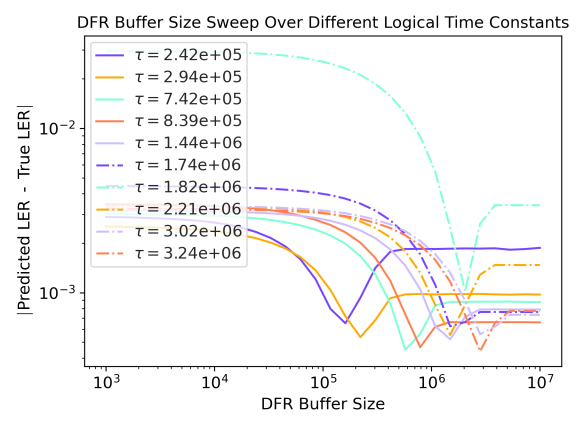}
    \caption{
        The absolute difference (L1 norm) between the predicted LER and the true LER, over logarithmic DFR buffer sizes for LER traces of several $d=3$ surface codes exhibiting varying magnitude of logical drift constant.
        Generally, increasing buffer size yields better prediction performance (lower L1-norm is better).
        However, each trace exhibits an inflection point; there exists an optimal buffer size where the L1-Norm begins to increase.
        Ideally, the optimal buffer size should increase with LDC; however, this is only true for the latter two traces (orange and light purple), and the former 3 traces exhibiting lower optimal buffer sizes with higher LDC instead.
        This may be a result of using only the mean of the temporal DFR window, rather than a more nuanced method, and leaves an interesting avenue available for future work.}
    \label{fig:dfr_buffer_expt}
\end{figure}

In practice, drift rates of hardware parameters are inaccessible (though we do expect any emergent patterns could be learned).
Instead, during computation we will only have access to a history of DFR information (derived from the syndromes we measure).
If we use only a single DFR value, e.g.
the most recently observed DFR, this can result in a warped view of the logical error rate for several reasons: 
\begin{enumerate}
    \item While there is a strong correlation between DFR and logical error rate, there can be variance between sequential measurements and LER is an emergent property.
Individual DFR can correspond to a measurement round from any number of tiles with widely varying logical error rates (e.g.
up to an order of magnitude).
    \item The trivial detector (all 0s, i.e.
DFR = 0) appears frequently and non-trivially even for moderate logical error rates and therefore is impossible to get any meaningful information about current tile performance.

\end{enumerate} 
A good prediction method should instead use a collection of recent DFR values to decipher the current logical error rate.
We consider a buffer of size $k$ which stores the $k$ most recent DFRs of every tile individually.
We query the predictor using the mean DFR over this range, and by default use the median value of the confidence interval, though we study the choice of confidence interval in Figure \ref{fig:confidence_intervals}, which is important when we care about balancing both accuracy and safety.
In Figure~\ref{fig:dfr_buffer_expt}, we study the choice $k$ for various drift speeds in the consistent drift model.
Importantly, small buffer sizes fluctuate rapidly in their predicted logical error rate, which illustrates the point from above.
Highly volatile predictions can be safe, but will result in excess resource overheads, regardless of the choice of response.
However, large buffer sizes shift too slowly to accurately keep up with the true drift rate.

At the systems level, we care not only about a single tile, but every possible logical patch.
In this case, we should keep track of these buffers for every location, even those which are not occupied by any information so that we can maintain a running list of "acceptable" locations in the system; this is especially relevant in our proposed relocation response when we should only move to tiles which are predicted to be good for a nontrivial number of cycles.

\subsection{Detection}
\begin{figure}[h]
    \centering
    \includegraphics[width=0.45 \textwidth]{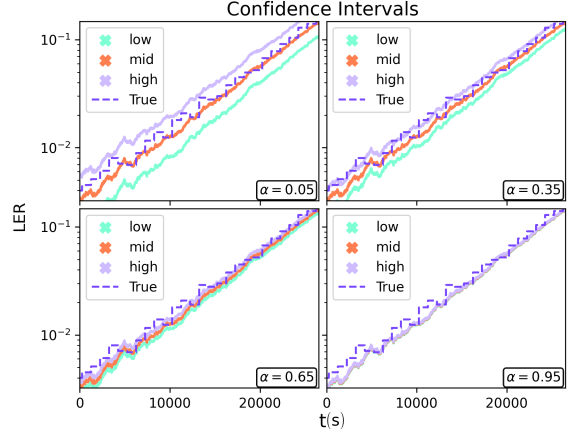}
    \caption{The fit to the DFR vs.
LER data yield a predictor that can estimate the LER for a given observed DFR.
We can tune the parameter $\alpha$ to determine the width of the prediction confidence interval, which allows us to tune the sensitivity of the drift detection module.
A lower value of $\alpha$ yields a larger confidence interval, making the detection module more sensitive.}
    \label{fig:confidence_intervals}
\end{figure}

We can use the DFR vs.
LER fits discussed in Figure \ref{fig:dfr-vs-ler} to effectively predict the LER of a logical qubit based on the recent history of its stabilizer measurements.
We consider a detection method parameterized by a temporal window size, which determines how many prior detection events are averaged together into a DFR, and a confidence parameter $\alpha$, which determines the size of the confidence interval by scaling the fit parameter uncertainty by the corresponding z-score (lower $\alpha$ leads to a larger confidence interval of possible LER values - see Figure~\ref{fig:confidence_intervals}).

\subsection{Prediction}
Though DFRs can be used individually to predict LER, as with most computer systems, maintaining a history of prior values can lead to more accurate predictions (in the same way branch prediction does).
Predicting LER from DFR is no different; maintaining a window/buffer of recent DFRs can help infer the current LER of a surface code patch more accurately than a single DFR alone.
Moreover, as we increase the number of available historical values, the higher the accuracy with which we can predict this LER (see Figure~\ref{fig:dfr_buffer_expt}).

That is, generally larger DFR buffers tend to yield more accurate LER predictions, allowing for more informed dynamic decisions at runtime (so long as the drift is slow enough).
Though this comes at a price; controllers need to be able to handle large buffers per logical patch to fully take advantage of the increased prediction accuracy.
Note that of course maintaining a larger DFR buffer also allows access to any buffer size beneath it.
Moreover, if drift is too fast, a larger DFR buffer may be too sluggish, or lazy, to reasonably predict the LER.
The sweet spot instead lies with a DFR buffer size that matches the speed of error drift.

The more closely the DFR buffer size matches the drift speed, the more ideal the predictor performance, minimizing the number of and length of breaches of the target LER (see Figure~\ref{fig:dfr_buffer_expt}).
\begin{figure}[ht]
    \centering
    \includegraphics[width=0.45\textwidth]{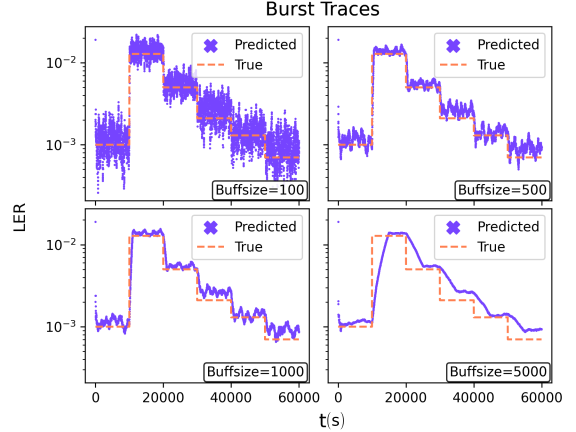}
    \caption{LER traces accompanied by predictions using DFR
buffers of various sizes. Predicting LER using historical DFRs
can be robust both under steady-state behavior, and under
unusual circumstances such as cosmic rays. Similar to static
drift, there exists an optimal DFR buffer size that i) yields
the most accurate LER predictions, and ii) offers the best
responsiveness to drift.}
    \label{fig:burst_traces}
\end{figure}
\subsubsection{Burst Errors}
While many studies characterize drift into a single, learnable parameter, in-practice drift parameters can deviate during program execution. One such extreme case is that of burst error, wherein error rates can drastically spike in a short amount of time. This can roughly be viewed as a temporally local, extremely low, logical drift constant. We therefore expect that the ideal DFR buffer size to be much smaller except during steady state behavior, where we should be less sensitive and therefore choose a much larger buffer. Specifically, in systems which experience both types of errors simultaneously, we need to balance being both responsive when bursts occur, while maintaining larger buffers to determine more aggregate drift behaviors dynamically.
This motivates a hierarchical choice of k, which examines multiple buffer sizes in tandem to make a prediction. This would be an interesting future research direction to improve the performance of the predictor, though we leave this to future work. In Figure~\ref{fig:burst_traces}, we demonstrate the performance of the predictor for several DFR buffer sizes undergoing a burst in physical error rates. 

\begin{figure}[ht]
    \centering
    \includegraphics[width=0.45\textwidth]{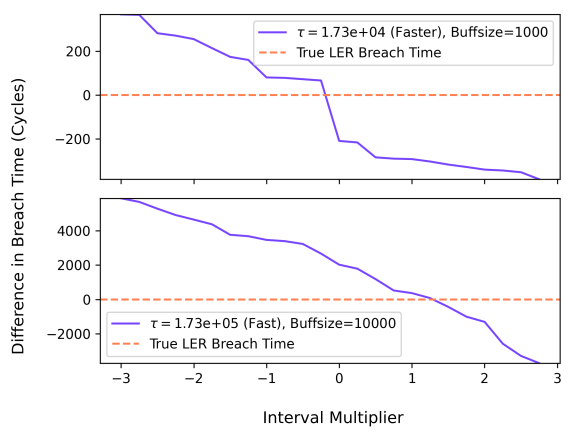}
    \caption{Prediction rarely results in detecting an LER breach (i.e.
surpassing target threshold) at the same time as the true LER.
Ideally, parameters should be set to minimize the time gap between the two.
The x-axis indicates the offset from the `best' preset. That is, the difference between the `high'/`low' interval bounds for a given $\alpha$, and the `best' (median) value for the interval. This difference is then multiplied by the `interval multiplier', and added to the `best' prediction value, to give the used prediction. 
The y-axis is then the relative gap between the true LER's breach time and the predictor's breach time, when using this interval multiplier.
The objective is to minimize this difference, i.e.
be as close to zero as possible.
Note that -1 corresponds to the `high' preset, while +1 corresponds to the `low' preset.}
    \label{fig:breach_diffs}
\end{figure}
\subsection{Prediction Timing}
Once a target LER threshold is set, the predictor module needs to be tuned as to prevent any LER breaches (surpassing the target LER).
Prediction will rarely detect an LER breach at the same time the true LER surpasses the target LER.
However, performing a response preemptively should take priority when tuning the predictor.
Responses take time to execute, and thus a marginal gap between true LER breach time and the predicted LER should exist.
Sweeping the confidence interval, we see the predictor module can be tuned to be zealous or sluggish in this regard (see Figure.~\ref{fig:breach_diffs}).
The difference from the detected breach time and the real breach time grows as we increase the magnitude of the confidence interval.
Positive differences indicate zealous predictions and are thus the prediction breaches the target LER ahead of when the true breach occurs, whereas negative differences indicate the prediction breaches after the true breach occurs.

\begin{figure}[!h]
    \centering
\scalebox{0.8}{
    \includegraphics[width=0.45\textwidth]{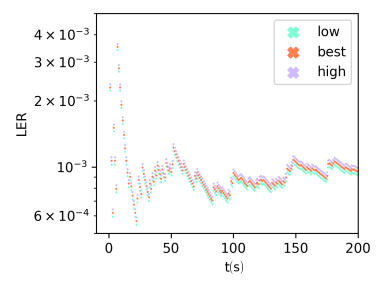}}
    \caption{Warm-up behavior of a $d=3$ surface code trace when using a DFR buffer size of 1000.
Due to the sudden change in mean DFR upon start-up, prediction can tend to sporadic values.
Consequently, a warm-up period is required prior to any meaningful prediction performance.}
    \label{fig:warm-up}
\end{figure}
\subsection{Warm-Up}
Similar to branch predictors in classical computer architecture, the DFR buffer is empty at the beginning of program execution.
As the program executes, the buffer saturates.
During a cold-start, the prediction accuracy can vary greatly due in part to i) the potential volatility of the qubit drift rates and ii) the increasing size of the DFR buffer as the warm-up period occurs (more DFRs populate the buffer with each cycle).
Thus, it's important to include a warm-up period following any recalibration period to obtain ideal performance from each tile.
During this warm-up period, we perform quantum error correction on a dummy logical qubit.
While no corrective actions are taken, the stabilizer constructions are executed each round each cycle such that the detectors are able to fire.
No data is maintained on the data qubits, but because the stabilizer circuits are being executed as though error correction was occurring, detector fire rates are able to be captured via error detection.
This informs our LER predictor without incurring any additional decoder overhead; the syndrome is solely extracted to compute the DFR (and subsequently approximate the LER).

\section{Response and Calibration}
Given this robust and reliable LER prediction method, appropriate action must be taken in the event an LER is predicted to breach the target LER.
Otherwise, continuing execution runs the risk of having to restart computation due to erroneous operations. In this section, we describe implementations and tradeoffs for potential calibration responses once a threshold breach has been detected.
We propose remapping as an efficient in-situ recalibration method, which relocates logical qubits to other logical patches in the architecture upon detecting a breach.
We then compare remapping to super-stabilizer code deformations as a baseline, and supplement with numerical studies to determine under which regimes each method performs better.
\begin{figure}[h]
    \centering
    \includegraphics[width=0.45\textwidth]{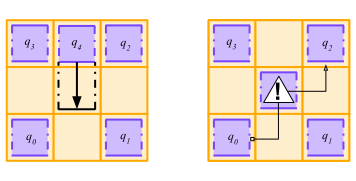}
    \caption{
    An example of a naive remap operation, which we avoid.
    \textbf{Left:} $q_4$ scheduling a remap into a routing channel. 
    \textbf{Right:} The instruction $\text{CNOT}\ q_0,\ q_2$ now cannot execute. This is an example of a i) a qubit being cordoned off, wherein the only fix is scheduling another remap operation for $q_4$ and ii) general operation congestion has to be rerouted to a different routing channel. (were the architecture larger), exacerbating execution times.}
    \label{fig:comm_issues}
\end{figure}
\subsection{Remapping}
Upon detecting an LER breach, a remap operation is triggered.
Remaps aim to i) \textit{dynamically} shift a logical qubit to a different patch to satisfy a LER target and ii) provide drifted logical patches a chance to recalibrate before incurring additional retry risk caused by drifted error rates. 
Once relocated, the logical qubit in question is situated on a logical patch at or below target LER. 
If no such patch is available, the lowest LER patch available is identified and remap occurs onto that.
Remaps are treated as all other program instructions, making them i) subject to routing constraints, and ii) simultaneously executable provided resources are available to facilitate them (e.g. routing space and space to remap to). 

The triggering patch is then scheduled for recalibration, disabling it for further use until recalibration completes.
Recalibration includes a warm-up period after completing, during which the tile remains disabled to allow the DFR buffer to saturate.
Specifically, the warm-up period's length is determined by however long it takes to saturate the DFR buffer.
The DFRs exhibited post-recalibration will be ideal since gate fidelities are strongest immediately post calibration. Intuitively, this should correspond to the best performing predictions. These immediate post-recalibration DFRs could potentially inform future predictions and/or detect drift due to DFR deviation from said ideal DFRs, and a follow-up study is required to explore this topic further.

Remaps that occur into routing space typically incur large program execution time overhead, as well as classical computational overhead on the controller.
Specifically, remaps onto routing space i) often decrease the number of operations routable in a single cycle and ii) may cordon off routing targets, requiring additional remap operations to make the operation possible in the first place (i.e. components become disconnected by series of adjacent logical qubits, see Figure~\ref{fig:comm_issues}).  

To ease routing complexity and controller overheads, qubits are instead remapped into buffers of dedicated surface code patches, called reloqation patches.
These buffers may reside anywhere in the architecture, and logical qubits may spend as much time as required before relocating to a different position (on the basis of LER target).
If a reloqation patch also experiences a threshold breach, another remap operation is scheduled onto another reloqation patch and subsequently marks the erring tile for recalibration, disabling it for further use until complete. 

Ideally, reloqation patches should be both i) uniformly distributed throughout the hardware to be reasonably accessible by all logical qubits mapped onto the architecture, and ii) accessible from as many routing channels as possible.
\begin{figure}[ht]
    \centering
    \includegraphics[width=\linewidth]{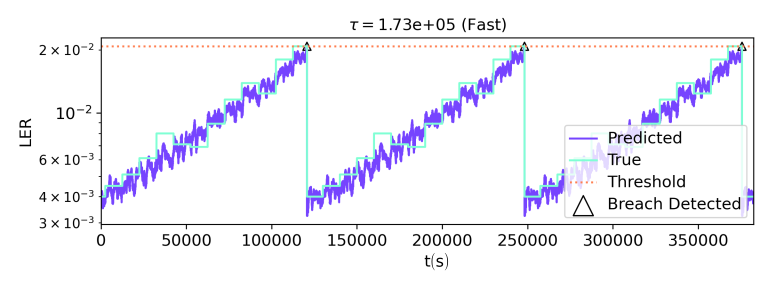}
    \caption{LER history of a logical qubit undergoing an architectural memory experiment. The predictor detects when a breach is imminent, and a remap is performed in response. Each drop in LER is indicative of a remap operation executing.}
    \label{fig:qtrace}
\end{figure}
\begin{figure}[ht]
    \centering
    \includegraphics[width=0.45\textwidth]{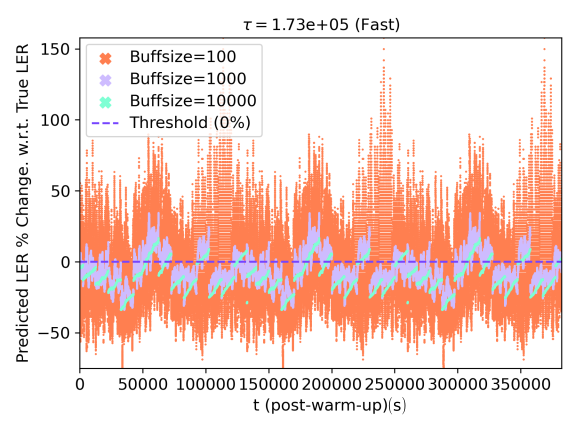}
    \caption{Relative percent error of an architectural memory experiment. The logical qubit is maintained on a $2\times2$ grid of logical tiles, wherein each tile shares a trace, though begins the experiment at a different point in the trace, ensuring there always exists a minimum LER tile. Similar to previous data, larger buffer sizes yield better prediction performance, even during remap operations. Proximity to zero is ideal.}
    \label{fig:remap_buffersize_test}
\end{figure}
\subsubsection{Remapping Example}
Using a $2\times 2$ tile architecture, we map a logical qubit onto a single tile, and give each tile the same LER+DFR trace. However, each tile is indexed at different points in the trace, with each being roughly a quarter of the total trace cycles apart from one another. This guarantees that one of the tiles exhibits a global minimum LER in the architecture. We then perform an expanded memory experiment on this demo architecture. Our predictor module is tuned with $\alpha=0.9$, and using the `low' preset. This results in an eager LER prediction that is relatively conservative. In doing so, we see that i) we're able to maintain high prediction performance while engaging our responses and ii) maintain a tolerable LER without ever exceeding our target threshold (see Figures~\ref{fig:qtrace} and~\ref{fig:remap_buffersize_test}). Note that all four logical patches were considered reloqation tiles for this experiment, allowing for remap to occur onto any tile. Moreover, recalibration takes roughly 250k cycles after detecting a breach, and remaps did not occur onto tiles undergoing recalibration.
\begin{figure}[ht]
    \centering
    \includegraphics[width=0.45\textwidth]{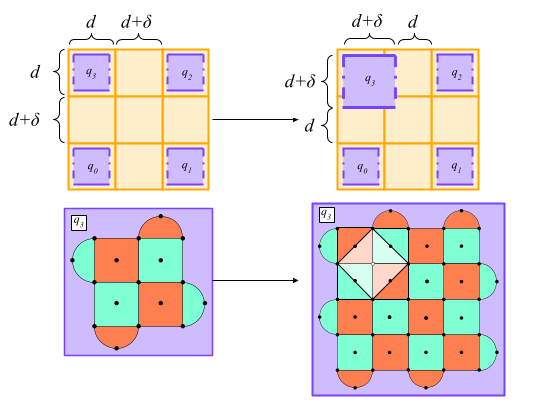}
    \caption{Demonstration of a code deformation response. Routing channels are maintained at a code distance $d+\delta$, while each logical qubit idles with distance $d < d+\delta$. Upon calibration, the logical qubit expands into the $\delta$ distance of the routing channels, and forms a superstabilizer around qubits in the first calibration group via code deformation. Here, a nominally $d=3$ qubit begins calibration, expanding into the routing space such that $q_3$ is now a $d'=5$, meaning $\delta=2$. After expansion (i.e. beginning calibration), operations are still able to use the remaining $d$-wide routing channels, and the expanded $q_3$ is still able to perform computation.}
    \label{fig:deform}
\end{figure}
\subsection{Code Deformation}
Here, we'll describe a recalibration schedule implemented via code deformation, inspired by~\cite{caliscalpel,surf_deform}, that recalibrates on a statically determined schedule, computed via device characterization prior to runtime.
Rather than maintain dedicated tiles to remap to, routing channels in the architecture are maintained at a distance $d+\delta$.
Under a sparse architecture, or architecture in which logical qubits are placed in a grid-like fashion with at least a single channel of ancilla space between each other, logical qubits are afforded the ability to expand into routing space due to the additional $\delta$ in code distance of the routing space (see Figure~\ref{fig:deform}).
Recalibration is then performed via statically scheduled independent subsets of physical qubits.
By deforming the code around each subset one at a time, and creating super-stabilizers~\cite{lin_codesign_2024} around the resulting holes in the topology, the logical qubit encoded on the logical tile remains fully operational/interactable for computation.
This process of deformation is repeated until all qubits in the tile have been recalibrated, thereby recalibrating the entire tile.
Formally, if $Q$ is the set of qubits in a given tile, let there be $m$ physical qubit calibration partitions, $s_k\subset Q$, such that $\mathcal{S}=\{s_k\}_{k=1}^{m},\ s_i\bigcap s_j=\varnothing$ for $i \neq j$, and $\bigcup_{k=1}^ms_k=Q$.
Once complete, the logical qubit contracts back to it's original code distance.
Note that $\delta $ may be tuned to provide as much additional code distance during expansion as needed when calibrating these independent subsets, but the effective distance should be approximately equivalent to that of the original distance in both the $Z$ and $X$ observables when in the expanded distance and under super-stabilizers induced by the deformation.
\begin{figure}[ht]
    \centering
    \includegraphics[width=0.45\textwidth]{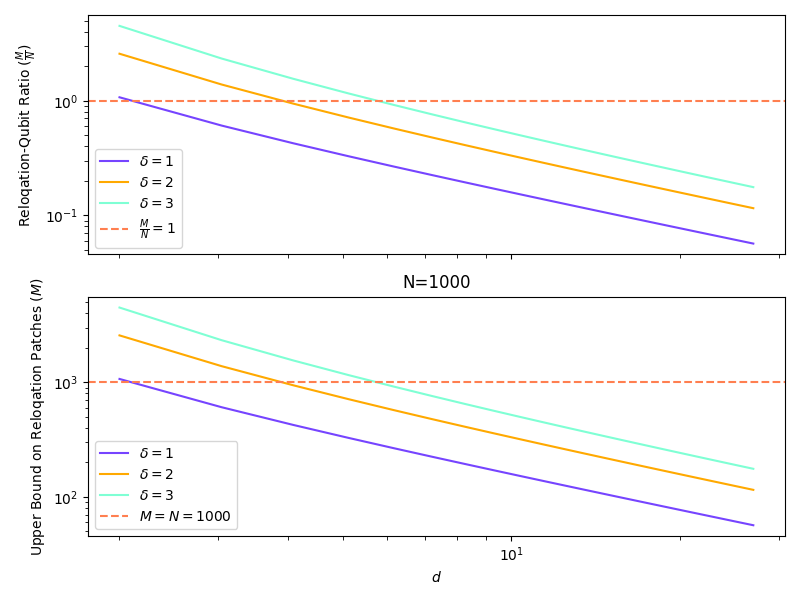}
    \caption{Crossover reloqation-qubit ratio for varying values of $d$ and $\delta$. The spatial efficiency of reloqation diminishes with $d^2$, making it more ideal for smaller distances, as the fraction of available reloqation patches while still remaining spatially efficient. We draw the same plot but solving for $M$ and using $N=1000$ qubits to demonstrate the actual number of reloqation patches that would be available at varying distances as well.}
    \label{fig:reloqation-qubit-crossover}

\end{figure}
\subsection{Spatial Tradeoffs}
For smaller distances, remapping demands a smaller spatial overhead than code deformation. Let us define a $2\times 2$ grid of logical tiles as a unit tile. One logical qubit will occupy the top left logical tile, while the rest of the tiles are dedicated to routing  (with the expectation that unit tiles would be placed adjacent to one another). Thus, a unit tile requires (noting that a single surface code tile requires $2d^2-1$ physical qubits) $8d^2-4$ physical qubits. 

A collection of unit tiles laid out in a grid is a good approximation of a sparse architecture. Under deformation calibration scheduling, however, routing tiles require $2(d+\delta )^2-1$ physical qubits. Thus, a unit tile under deformation recalibration scheduling requires 
\begin{align*}
3\cdot (2(d+\delta)^2-1) + (2d^2-1)\\
=8d^2+12d\delta  + 6\delta^2 - 4
\end{align*}
physical qubits. If we require $N$ logical qubits, and instead use $M$ dedicated reloqation patches for recalibration, we assign each logical qubit and reloqation tile onto a unit tile. The number of physical qubits required by each response is thus:
\begin{gather*}
    \text{Reloqation: } (N+M)\cdot4(2d^2-1)\\
    \text{Deformation: } (N) \cdot (8d^2+12d\delta  + 6\delta^2 - 4)\\
    \boxed{\implies\frac M N = \frac {3\delta} {4d^2-2}(2d+\delta).}
\end{gather*}
We also compute the reloqation-qubit ratio $\frac M N$. For remapping, we'll define spatial efficiency as the number of reloqation tiles not exceeding the number of logical qubits. This manifests in $ \frac {3\delta} {4d^2-2}(2d+\delta) = \frac M N \leq 1$. However, note that this spatial efficiency diminishes with $d^2$. That is, with larger distances, the number of reloqation patches we can use while still still being spatially efficient decreases (see Figure~\ref{fig:reloqation-qubit-crossover}). This makes sense as variance in hardware error rates also has diminishing impact as code distance becomes sufficiently larger, Moreover, $\delta$ can be adjusted to accommodate larger recalibration subsets, resulting in faster overall recalibration due to the additional parallelization at the cost of additional space.

\begin{figure*}[ht!]
    \centering
    \includegraphics[width=0.75\textwidth]{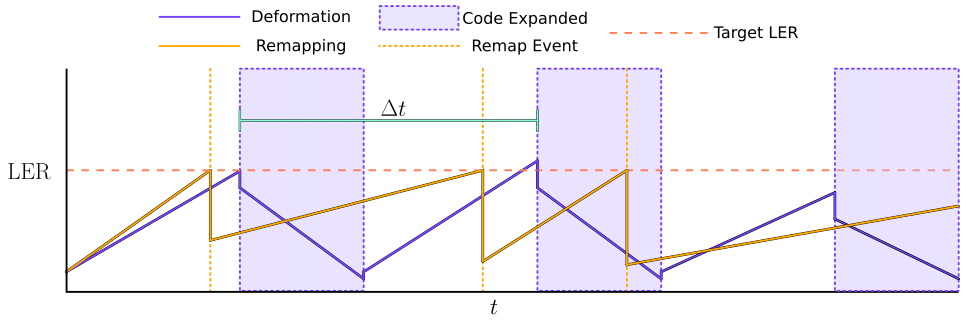}
    \caption{\emph{Conceptual} timing diagram for a system implementing dynamic remapping as a calibration scheduler against a system implementing deformations via static calibration scheduling with calibration frequency $f_{deform} = \frac 1 {\Delta t}$. $\Delta t$ is typically determined via characterization prior to execution such that the target LER is maintained or beaten. Note that, because deviations in drift parameters can occur mid-execution, static scheduling may i) recalibrate too late, as in the second chronological code expansion where the target LER is breached, or ii) recalibrate too early, meaning more time will have been spent in an expanded state than required, as in the third chronological code expansion.}
    \label{fig:deformation_demo}
\end{figure*}
\subsection{Runtime Tradeoffs}

Regardless of response, it is important to choose a well thought out target LER, as it directly impacts how often a response is executed.
For remapping, more movement operations will occur the lower the target. 
For deformation, the frequency that expansion occurs at will increase, and consequently the time spent in an expanded code distance. 

Moreover, a poorly chosen target LER can be detrimental to program execution.
Too loose/high, calibration will be done lazily, incurring additional retry risk to program execution. 
Conversely, in both cases, stringent LER targets lead to varying execution detriment.
For remapping, excessive LER constraints will cause thrashing.
Movement operations will enqueue as each tile is deemed unusable and/or undergoing recalibration.
Deformation instead exhibits unwieldy portions of program execution in an expanded state, causing additional overheads as expansions hog resources.

Deformation, while not spatially bounded in the number of possible calibrations executing simultaneously at once in higher distance codes, relies on prior characterization of logical patches in order to generate a deformation/calibration schedule. Yet, in practice, drift is not characterized by a single parameter, despite it's usage in simulations. Statically scheduled calibrations can suffer from deviations in drift parameters, causing calibration to begin either too early or too late. In the case of the former, the portion of time spent in an expanded distance is increased, further exacerbating resource constraints. Conversely, the latter can cause execution to incur additional retry risk since operational LER may breach the target LER.

In these two regards, it's generally advantageous to be dynamic. Rather than using predetermined drift constants to schedule calibrations, remapping relies on the LER prediction to dictate whether to trigger a movement operation and begin recalibration on the erring tile. 
While reliant on the performance of the predictor, this allows it to adjust to potential fluctuations in drift parameters during runtime, without additional characterization. 
Moreover, we've shown the performance of the predictor to be capable when using DFR sampled from real hardware (See Figure~\ref{fig:drift-examples}). 
Prediction accuracy can also be increased as training data becomes more intricate and prediction models adjust accordingly, as seen in~\cite{xiao_lukin_2026,zheng_liang_2026,remm_wallraff_2025}.

As with most engineering problems, a hybrid solution is ideal to obtain the benefits of both response methods.
Deformation should be made dynamic and used in the case of larger codes, while smaller codes should employ dynamic remapping.
In both cases, a target LER should be set, and an LER predictor should be used to determine when to trigger a response. 
Current deformation+calibration schemes~\cite{caliscalpel,surf_deform} instead rely on static characterization prior to runtime. 
When combined with an LER predictor, deformation and consequently recalibration can be done dynamically on an as-needed basis.
This could be another direction of future research. 
Deformation could be done to calibrate individual physical qubits dynamically, without the need to execute all calibration groups sequentially after the predictor triggers a response.

\section{Conclusion}
Error rates can vary both spatially and temporally on real quantum hardware, which many current analyses fail to account for. 
To exacerbate this issue, characterization and calibration information tends to quickly become stale as a result, yielding low fidelity operations and leaving logical qubits enduring excessive error rates. 
In this work, we draw a clear connection between detector fire rate, a readily available metric on surface codes, and logical error rate, using circuits generated with error rates of real-world device data. 
This connection is extrapolated to a fit, which drives a logical error rate predictor module for the surface code, and is extendable to other stabilizer codes. 
We then show that this predictor can be tuned to be performant under static drift models. 

We go on to present a recalibration method, remapping, which uses dedicated tiles in a surface code architecture for relocating logical qubits when their current tile requires recalibration, as dictated by an logical error rate predictor. Recalibration is then performed on any tile that is remapped from.
We show that this technique is spatially efficient for a reloqation-qubit ratio that shrinks with code distance.
More specifically, the ratio of relocation tiles to logical qubits shrinks as we increase code distance, but is approximately upper bounded by the number of logical qubits (N) for small code distances. 
We then compare this to current in-situ recalibration methods, which use code deformation to form super-stabilizers around physical qubits to calibrate them while keeping logical tiles (and any existing logical qubits) functional for use during computation. 
However, these methods rely on static analysis of drift parameters prior to runtime to generate a calibration schedule, even though these drift parameters can change during runtime, causing recalibration to occur either too early or too late.
Using statically determined parameters can thus result in i) additional resource overheads caused by larger portions of execution spent in an expanded code distance when calibrating too early, or ii) incurring additional retry risk to program execution as recalibration is done too late. 
Ultimately, a hybrid solution of dynamic remapping for smaller distances and dynamic deformation for larger distances is required to obtain the benefits of both methods at efficient overheads. 
Both methods should be based on dynamic LER prediction to allow for as-needed recalibrations rather than static recalibration scheduling, which can subject program execution to additional retry risk and runtime overheads.

\bibliographystyle{acm}
\bibliography{refs}

\end{document}